\documentclass[10pt,conference]{IEEEconf}  
\renewcommand{\baselinestretch}{0.95}
\usepackage{amsmath}
\usepackage{amsfonts}
\usepackage{graphicx}
\usepackage{float}
\usepackage{color}
\graphicspath{{Figures/}}

\newcommand{\cmt}[1]{} 

\usepackage{units} 

\usepackage{amsthm}					
\newcounter{prb}
\newtheorem{prob}[prb]{Problem} 	

\usepackage{pdflscape} 	
\usepackage{soul} 		

\usepackage{cite}  		

\usepackage{tikz}
\usetikzlibrary{shapes,arrows,positioning,shadows}
\usetikzlibrary{calc}

\usepackage{lipsum} 
\usepackage{lettrine} 
\usepackage{placeins} 

\usepackage{array}

\usepackage{cancel}

\usepackage{mathtools}

\usepackage{cuted}

\usepackage{multirow} 

\usepackage{algorithm2e} 
\RestyleAlgo{ruled} 

\usepackage[final]{pdfpages} 

\IEEEoverridecommandlockouts
\usepackage{acro}

\DeclareAcronym{fgt}{
	short = FGT,
	long  = fixed-gear transmission
}
\DeclareAcronym{mgt}{
	short = MGT,
	long  = multiple-gear transmission
}

\DeclareAcronym{cvt}{
	short = CVT,
	long  = continuously variable transmission
}

\DeclareAcronym{pmp}{
	short = PMP,
	long  = Pontryagin’s Minimum Principle
}

\DeclareAcronym{miocp}{
	short = MIOCP,
	long  = mixed-integer optimal control problem
}

\DeclareAcronym{em}{
	short = EM,
	long  = electric motor
}
\DeclareAcronym{rmse}{
	short = RMSE,
	long  = root-mean-square error
}
\DeclareAcronym{socc}{
	short = SOCC,
	long  = second-order cone constraint
}

\DeclareAcronym{cp}{
short = COP,
long  = continuous optimization problem
}

\DeclareAcronym{gp}{
	short = GOP,
	long  = gearshift optimization problem
}

\DeclareAcronym{2gt}{
	short = 2GT,
	long  = 2-speed MGT
}

\DeclareAcronym{3gt}{
short = 3GT,
long  = 3-speed MGT
}

\DeclareAcronym{4gt}{
short = 4GT,
long  = 4-speed MGT
}

\title{Time-optimal Design and Control of Electric Race Cars Equipped with Multi-speed Transmissions}
\author{Camiel Cartignij and Mauro Salazar
	\thanks{The authors are with the Control Systems Technology section, Department of Mechanical Engineering, Eindhoven University of Technology (TU/e), Eindhoven, 5600 MB, The Netherlands.
		E-mails: {\tt\footnotesize c.j.g.cartignij@student.tue.nl, m.r.u.salazar@tue.nl}} 
}
\begin{document}

\maketitle

\begin{abstract}
	This paper presents a framework to jointly optimize the design and control of an electric race car equipped with a \ac{mgt}, specifically accounting for discrete gearshift dynamics.
We formulate the problem as a \ac{miocp}, and deal with its complexity by combining convex optimization and \ac{pmp} in a computationally efficient iterative algorithm satisfying necessary conditions for optimality upon convergence.
Finally, we leverage our framework to compute the achievable lap time of a race car equipped with a \ac{fgt}, a \ac{cvt} and an \ac{mgt} with 2 to 4 speeds, revealing that an \ac{mgt} can strike the best trade-off in terms of electric motor control, and transmission weight and efficiency, ultimately yielding the overall best lap time.

\end{abstract}
\acresetall

\section{Introduction}
\lettrine{T}{he} electrification of automotive powertrains has sparked significant attention over the last years. Conventional passenger vehicles are being replaced by hybrid and fully electric vehicles \cite{IEA2021}, while existing racing classes are also hybridized, and new fully electric racing classes are emerging. 
In motorsport, every millisecond counts, so all involved technologies are pushed to their limits. For electric racing, the available battery energy is a significant limitation, which means that optimal energy management and \ac{em} efficiency can make the difference to win a race. 
To achieve maximum performance while keeping the \ac{em} within its most efficient operating range, various transmission types can be used, which yield control over the \ac{em} operating range, at the cost of the transmission's own efficiency and weight. 
At one end of the spectrum, the \ac{fgt} is light and efficient, while providing little control over the operating range of the \ac{em}. At the other end, the \ac{cvt} provides continuous control over the \ac{em} operation, but at the cost of increased weight and a lower efficiency. Between these extremes, a \ac{mgt} could balance the efficiency of an \ac{fgt} with the operating range control of a \ac{cvt}. 
To select the optimal transmission for any given application, a joint optimization of transmission design and \mbox{powertrain} control is required. However, the gearshifts of an \ac{mgt} introduce discrete dynamics, which turns the optimization of an \ac{mgt}-equipped race car into a \ac{miocp}. \ac{miocp}s are generally $\mathcal{NP}$-hard, and can be very difficult to solve \cite{Sager2005}.
Against this backdrop, this paper presents a computationally efficient algorithm for the optimization of the design and control of an \ac{mgt}-equipped electric race car.

\begin{figure}[t]
	\def\nodeSize{8.3mm}
\def\smallNodeSize{.5*\nodeSize}
\tikzstyle{blockAlg} = [draw, rectangle, minimum height=\nodeSize, minimum width=\nodeSize, align=center, thick]
\tikzstyle{blockBkg} = [draw, rectangle,drop shadow,top color=white, minimum height=\nodeSize, minimum width=\nodeSize, align=center, thick]
\tikzstyle{blockSmall} = [draw, rectangle, minimum height=\smallNodeSize, minimum width=\nodeSize, align=center, thick]

\tikzset{
	partial ellipse/.style args={#1:#2:#3}{
		insert path={+ (#1:#3) arc (#1:#2:#3)}
	}
}

\pgfdeclarelayer{bg}    
\pgfsetlayers{bg,main}  
\def\distanceNodes{8mm}
\def\spacer{0.54em}
\pgfmathsetlengthmacro\frameSize{(.5*\nodeSize+\spacer)}
\pgfmathsetlengthmacro\smallFrameSize{(.5*\smallNodeSize+\spacer)}
\begin{footnotesize}
	\begin{tikzpicture}[scale=1, every node/.append style={outer sep=0pt}, >=stealth,	]
		\tikzstyle{every node}=[font=\footnotesize]

		\node[blockBkg] at (current page.north west) (es) {\parbox[b][2em]{2em}{\centering BAT}};
		\node[draw,circle,anchor=west] at ($(es.east)+(\distanceNodes,0)$) (sum1) {};
		\node[blockAlg,anchor=west] at ($(sum1.east)+(\distanceNodes,0)$) (inv) {\parbox[b][2em]{2em}{\centering INV}};
		\node[blockAlg,anchor=west] at ($(inv.east)+(\distanceNodes,0)$) (mot) {\parbox[b][2em]{2em}{\centering EM}};
		\node[blockAlg,anchor=west] at ($(mot.east)+(\distanceNodes,0)$) (trans) {\parbox[b][2em]{2em}{\centering GB}};
		\node[blockBkg,anchor=west] at ($(trans.east)+(\distanceNodes,0)$) (wheel) {\parbox[b][2em]{2em}{\centering W}};
		
		\node[blockSmall,anchor=north west] at ($(inv.south west)+(0,-2*\spacer)$) (fgt) {FGT};
		\node[blockSmall] at ($(fgt)+(0,-\frameSize)$) (mgt) {MGT};
		\node[blockSmall] at ($(mgt)+(0,-\frameSize)$) (cvt) {CVT};
		
		\draw[<->] (es) -- (sum1) node[midway,above] {\parbox[t][2.7mm]{\distanceNodes}{\centering $P_\mathrm{b}$}};
		\draw[<->] (sum1) -- (inv) node[midway,above] {\parbox[t][2.7mm]{\distanceNodes}{\centering $P_\mathrm{dc}$}};
		\draw[->] (sum1) --++(0,-1.5*\distanceNodes) node[below] {\parbox[t][2.7mm]{\distanceNodes}{\centering $P_\mathrm{aux}$}};
		\draw[<->] (inv) -- (mot) node[midway,above] {\parbox[t][2.7mm]{\distanceNodes}{\centering $P_\mathrm{ac}$}};
		\draw[<->] (mot) -- (trans) node[midway,above] {\parbox[t][2.7mm]{\distanceNodes}{\centering $P_\mathrm{m}$}};
		\draw[<->] (trans) -- (wheel) node[midway,above] {\parbox[t][2.7mm]{\distanceNodes}{\centering $P_\mathrm{gb}$}};
		
		\begin{pgfonlayer}{bg}
		\draw[dashed,line width=1.5,color=black!20] 
		($(cvt.south west)  +(-\spacer,-\spacer)$) -| 
		($(trans.north east)+( \spacer, \spacer)$) --
		($(trans.north west)+(-\spacer, \spacer)$) |- 
		($(fgt.north west)  +(-\spacer, \spacer)$) -| 
		($(cvt.south west)  +(-\spacer,-\spacer)$);
		\end{pgfonlayer}
		\node[text width=2.6cm,anchor=west] at ($(fgt.east)+(.25*\distanceNodes,0)$) (fgt1) {$\gamma=\gamma_1$};
		\node[text width=2.6cm,anchor=west] at ($(mgt.east)+(.25*\distanceNodes,0)$) (mgt1)  {$\gamma\in\{\gamma_1,\dots,\gamma_{n_\mathrm{gear}}\}$};
		\node[text width=2.6cm,anchor=west] at ($(cvt.east)+(.25*\distanceNodes,0)$) (cvt1) {$\gamma\in[\gamma_\mathrm{min},\gamma_\mathrm{max}]$};
		
		\draw (es.north east) [partial ellipse=270:180:{\nodeSize} and {0.5*\nodeSize}];
		\node[text width=\nodeSize,align=right] at ($(es.north)+(0em,-1.8mm)$) (bat1) {$E_\mathrm{bat}$};
		
		
		\draw (wheel.north east) [partial ellipse=270:180:{\nodeSize} and {0.5*\nodeSize}];
		\node[text width=\nodeSize,align=right] at ($(wheel.north)+(0em,-1.8mm)$)  {$E_\mathrm{kin}$};
		
	\end{tikzpicture}
\end{footnotesize}

\vspace{-0.9em}
	\caption{Schematic layout of the electric race car powertrain under consideration, consisting of a battery (BAT), an inverter (INV), an electric motor (EM), a transmission (GB) and the wheels (W). The gear-ratio $\gamma$ is a control input that must be chosen from a finite (FGT and MGT) or continuous (CVT) set, which is in turn subject to design. The arrows indicate power flows between components.}
	\label{fig:ptdiag}
\end{figure}
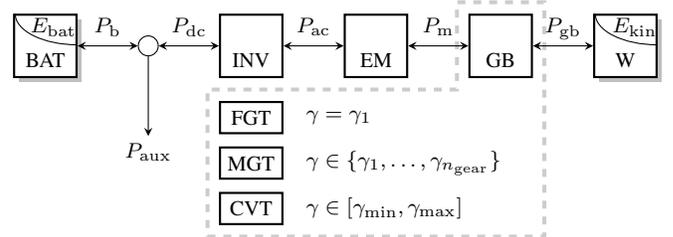

\textit{Related literature:} The mixed-integer time-optimal design and control problem studied in this paper pertains to two main research streams, namely, energy management and time-optimal control of (hybrid) electric vehicles.

For energy management, mainly non-causal optimization is applied on drive cycles with known velocity and torque requirements. Joint optimization of the gearshift strategy and gear ratio design has been carried out using dynamic programming \cite{GaoLiangEtAl2015}\cmt{401}, combined dynamic programming and convex optimization \cite{HurkSalazar2021}\cmt{103} and mixed-integer nonlinear optimization \cite{YuZhangEtAl2020}\cmt{531}. For the design and control of an \ac{fgt}- and \ac{cvt}-equipped vehicle, convex optimization is employed by~\cite{VerbruggenSalazarEtAl2019,HurkSalazar2021}.
However, since these methods use known drive cycles, none of them can be directly applied to time-optimal control problems, where the velocity trajectory is unknown. 

For the time-optimal control of race cars, \cite{EbbesenSalazarEtAl2018} proposes a distance-based convex optimization framework, which is extended to consider the design and control of an \ac{fgt} and \ac{cvt} in~\cite{BorsboomFahdzyanaEtAl2021}. In~\cite{DuhrChristodoulouEtAl2020}, the \ac{mgt} gearshift is optimized by using iterative dynamic programming with \ac{pmp} and convex optimization, while \cite{BalernaNeumannEtAl2021} uses outer convexification and nonlinear programming to optimize the gearshift. However, both methods lack optimality guarantees and solely focus on control strategies.

In conclusion, to the best of the authors' knowledge, there are no frameworks for time-optimal control of race cars considering both gearshifts and gear ratio design.

\textit{Statement of Contributions:} This paper provides a computationally efficient framework to optimize the design and control of the electric race car powertrain shown in Fig.~\ref{fig:ptdiag}. To this end, we introduce an efficient iterative algorithm to optimize the \ac{mgt}-equipped race car, combining convex optimization with \ac{pmp} to efficiently account for the discrete gearshift dynamics. Finally, we leverage our algorithm to compare the performance of a race car equipped with an \ac{fgt}, a \ac{cvt} and an \ac{mgt} with 2 to 4 speeds on the Zandvoort race track.

\section{\ac{mgt} Optimization Methodology}\label{sec:optimization}
In this section, we define the minimum-lap-time design and control problem for the powertrain shown in Fig.~\ref{fig:ptdiag}, together with algorithms to solve the optimization problems. 

\subsection{Optimization Problem Definition}
We base our optimization framework on the convex framework developed in \cite{BorsboomFahdzyanaEtAl2021} for the time-optimal design and control of an \ac{fgt}- and a \ac{cvt}-equipped race car, while we include an \ac{mgt} model, and the more accurate vehicle dynamics and inverter models from \cite{KampenHerrmannEtAl2023}. 
This provides us with a minimum-lap-time design and control problem in the space-domain, which grants us a finite problem horizon $S$, and allows us to optimize over the position on track $s$. For the sake of brevity, we do not show our component models here. 

Before defining our optimization problem, we divide our optimization variables as follows: 
First, we consider the design variables: $p=\gamma_1$ for the \ac{fgt}, $p=	\{\gamma_1,\dots,\gamma_{n_\mathrm{gear}}\}$ for the \ac{mgt} and $p=\gamma_\mathrm{max}$ for the \ac{cvt}. Here, $\gamma_i$ is the ratio of gear $i$, $n_\mathrm{gear}$ is the total number of \ac{mgt} gears, and $\gamma_\mathrm{max}$ is the maximum ratio for the \ac{cvt}. For notational convenience, we model the design variables as state variables with zero dynamics,
$$ \frac{\mathrm{d}\gamma_i(s)}{\mathrm{d}s} = 0, $$
which leaves only the initial conditions as free optimization variables. With this, we consider the state variables as
\mbox{$x= \{E_\mathrm{bat},E_\mathrm{kin},p\}$}, where $E_\mathrm{bat}$ and $E_\mathrm{kin}$ are the battery energy and kinetic energy of the vehicle, respectively. As control inputs we consider 
$u=	\{F_\mathrm{m}, F_\mathrm{brk,F}, F_\mathrm{brk,R}\}$, where $F_\mathrm{m}$ is the mechanical motor force, assuming a rear-wheel driven vehicle, and $F_{\mathrm{brk,F}}$ and $F_{\mathrm{brk,R}}$ are the brake forces at the front and rear axle, respectively. For the \ac{cvt} we further consider the gear ratio at a given position on track $\gamma(s)$ as an input variable, whilst for the \ac{mgt} we consider the additional input $b(s)\in \{0,1\}^{n_\mathrm{gear}}$, which is a binary variable stating the active gear at a given position on track. 
With this, we formulate our problem to minimize the lap time $T$ as follows: 
\begin{prob}[Minimum-lap-time problem]\label{prob:convex}
	The time-optimal design and control strategies are the solution of
	\begin{eqnarray*}
		&&\!\min_{x,u}    T, \\
		&&\begin{aligned}
			\textnormal{s.t. }	& \textnormal{State dynamics and limits},&&&&\\
			&	\textnormal{Vehicle dynamics and limits,}	 &&\\
			&	\textnormal{Component models.}	 &&
		\end{aligned}
	\end{eqnarray*}
\end{prob}
Since all models and variables for the \ac{fgt} and \ac{cvt} optimization problems are continuous and convex, Problem~\ref{prob:convex} can be solved for the globally optimal solution in polynomial time \cite{BoydVandenberghe2004} using commercially available solvers. Conversely, due to the binary gearshift variable $b$, the \ac{mgt} optimization problem is a \ac{miocp}, which is computationally demanding, and often even intractable, hence calling for an alternative solution scheme, which we outline in Section~\ref{sec:iterative} below.

\subsection{Iterative algorithm}\label{sec:iterative}

\begin{figure}[bt]
	\centering
	\def\nodeWidth{18em}
\def\nodeWidthSmall{4em}
\def\nodeHeight{2em}
\def\nodeHeightSmall{.5\nodeHeight}
\tikzstyle{blockAlg} = [draw, rectangle, minimum height=\nodeHeight, minimum width=\nodeWidth, align=center, inner sep=10pt, thick]
\tikzstyle{blockSquare} = [draw, rectangle, minimum height=\nodeHeightSmall, minimum width=\nodeHeightSmall, align=center, inner sep=10pt, thick]

\tikzset{
	partial ellipse/.style args={#1:#2:#3}{
		insert path={+ (#1:#3) arc (#1:#2:#3)}
	}
}

\usetikzlibrary{arrows.meta}
\def\offcenter{0.2*\nodeWidth}
\def\distanceNodes{2.1em}
\def\spacer{0.5*\distanceNodes}
\begin{tikzpicture}[scale=1, every node/.append style={outer sep=0pt}, >=stealth,	]
	
	\node[blockAlg,inner sep=4pt]    at (0,0)                          (cp)   {\parbox[c][\nodeHeight]{\nodeWidth}{\centering  Continuous Optimization \\ $(x^k,\lambda^k,c_\mathrm{C}^k) = \textnormal{\ac{cp}}(b^{k-1})$}};
	\node[blockSquare,anchor=north,inner sep=4pt] at ($(cp.south)+(0,-\distanceNodes)$) (da)  {\parbox[c][\nodeHeightSmall]{7.2em-8pt}{\centering  Damping}};
	\node[blockAlg   ,anchor=north,inner sep=4pt] at ($(da.south)+(0,-\distanceNodes)$) (gp)   {\parbox[c][\nodeHeight]{\nodeWidth}{\centering  Gearshift Optimization \\ $(b^k,c_\mathrm{G}^k) = \textnormal{\ac{gp}}(x^k,\tilde{\lambda}^k)$}};
	\node[blockSquare   ,anchor=north,inner sep=4pt] at ($(gp.south)+(0,-\distanceNodes)$) (ch)   {\parbox[c][\nodeHeightSmall]{\nodeWidth}{\centering  Converged?}};
	\node[blockSquare   ,anchor=south west,inner sep=4pt] at ($(cp.north west)+(0,\distanceNodes)$) (k1)    {\parbox[c][\nodeHeightSmall]{\nodeWidthSmall}{\centering  $k=k+1$}};	
	
	\node[blockSquare,anchor=south,inner sep=4pt] at ($(cp.north)+(0   ,\distanceNodes)$)      (k0)    {\parbox[c][\nodeHeightSmall]{\nodeWidthSmall}{\centering  $k=1$}};
	
	\draw[-{Latex[round]}] (k0) -- (cp) node[midway,left] {$b^0$};
	\draw[-{Latex[round]}] (cp) -- (da) node[midway,left] {$x^k,\lambda^k$};
	\draw[-{Latex[round]}] (da) -- (gp) node[midway,left] {${x}^k,\tilde{\lambda}^k$};
	\draw[-{Latex[round]}] (gp) -- (ch) node[midway,left] {$b^{k},c^k_\mathrm{G}$};

	\draw[-]  ($(ch.south west)+(\offcenter,0)$) -- 	($(ch.south west)+(\offcenter,-\distanceNodes)$) node[midway,left] {No};
	\draw[-{Latex[round]}]  (k1) -- (k1 |- cp.north) 			  node[midway,left] {$b^{k-1}$};
	
	\draw[-{Latex[round]}]  ($(ch.south west)+(\offcenter,0)$) -- 	($(ch.south west)+(\offcenter,-\distanceNodes)$) -- ($(ch.south west)+(-\spacer,-\distanceNodes)$) |- 	(k1.west); 
	
	\draw[-{Latex[round]}] ($(ch.south east)+(-\offcenter,0)$) -- 	($(ch.south east)+(-\offcenter,-\distanceNodes)$) node[below] {$x^\star,b^\star,c^\star$};	
	\draw[-]  ($(ch.south east)+(-\offcenter,0)$) -- 	($(ch.south east)+(-\offcenter,-\distanceNodes)$) node[midway,left] {Yes};

	\path  ($(cp.south east) + (\spacer,-\spacer)$) --
	($(gp.north east) + (\spacer,\spacer)$)	  node[midway,left] {$x^k,\lambda^k,c^k_\mathrm{C}$};

	\draw[-{Latex[round]}] ($(cp.south east) + (-\offcenter,0)$) --
	($(cp.south east) + (-\offcenter,-\spacer)$) -|
	($(ch.north east)  +(\spacer,\spacer)$) -| 
	($(ch.north east)  +(-\offcenter,0)$);

\end{tikzpicture}
	\caption{Iterative algorithm for the optimization of an electric race car equipped with an \ac{mgt}. Given an initial gearshift trajectory $b^0$, the continuous variables are optimized in the COP. The resulting state and costate trajectories $(x,\lambda)$ are then used to re-optimize the gearshift trajectory in the GOP. These steps are iterated until the results converge.} \label{fig:it_scheme} 
\end{figure}
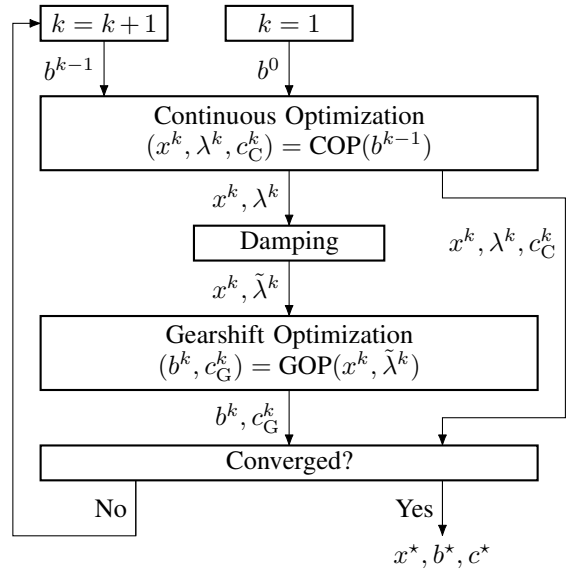

In this section, we describe the iterative algorithm, shown in Fig.~\ref{fig:it_scheme}, used to solve the optimal design and control problem of an \ac{mgt}-equipped race car.
We subdivide the \ac{miocp} into a \ac{cp}, which optimizes the continuous design and control variables for a given gearshift trajectory, and a \ac{gp}, which optimizes the gearshifts for a given state and costate trajectory.
We iterate between these problems until the results converge, using dampened costate trajectories to ensure convergence. 
To distinguish between iterations of our algorithm, we introduce the $(\cdot)^k$ notation to describe the current iteration $k\in \mathbb{N}$. To improve readability, we group our continuous inputs as $c=\{F_\mathrm{m}, F_\mathrm{brk,F}, F_\mathrm{brk,R}\}$. Finally, since the continuous inputs are optimized by both the \ac{cp} and \ac{gp}, we distinguish between the results of the respective problems using a $(\cdot)_\mathrm{C}$ and $(\cdot)_\mathrm{G}$ notation.

\subsection{Continuous Optimization Problem Definition} \label{sec:cop}
By providing a pre-determined trajectory $b^{k-1}(s)$, we can optimize our continuous variables using the convex \ac{cp}.
\begin{prob}[Continuous Optimization Problem]\label{prob:cp}
	We define the \ac{cp} for \ac{mgt} optimization as
	\begin{eqnarray*}
		&&(x^k,c^k_\mathrm{C})=  \arg\min_{x,c}     T,\\
		&&\begin{aligned}
			\textnormal{s.t.:}~	& \textnormal{State dynamics and limits,}&&&&\\
			&		\textnormal{Vehicle dynamics and limits,}	 &&\\
			&		\textnormal{Component models,}	 &&\\
			&				b(s)\coloneqq b^{k-1}(s)&&\forall s \in [0,S].&&
		\end{aligned}
	\end{eqnarray*}
\end{prob}
This fully convex \ac{cp} can be solved in polynomial time, similar to Problem~\ref{prob:convex} for the \ac{fgt} and \ac{cvt}. 

\subsection{Gearshift Optimization Problem Definition}\label{sec:gop}
We optimize the binary gearshift trajectory by applying \ac{pmp}, which allows us to efficiently solve the problem for each position on track independently. According to \ac{pmp} \cite{Bertsekas1995}, the optimal solution satisfies
$$	(u^\star) = \arg \underset{u}{\min}~\mathcal{H}(x^\star,\lambda,u), \label{eq:pmp}$$
where $(\cdot)^\star$ denotes an optimal trajectory, $\mathcal{H}$ is the Hamiltonian, and $\lambda$ represents the costates.
If we provide a pre-determined state and costate trajectory $(x,\lambda)$, the Hamiltonian minimization problem can be solved at every position on track independently, thereby significantly reducing the computational complexity. By minimizing the Hamiltonian for each gear option separately with convex optimization, the optimal trajectory can be determined by selecting the Hamiltonian with the minimum value at each position, which can be done in polynomial time.  
By providing a pre-determined state and costate trajectory $(x^{k}(s),\lambda^{k}(s))$, and by omitting the state dynamics and limits, we formulate the \ac{gp} as follows:

\begin{prob}[Gearshift Optimization Problem]\label{prob:gp}
	We define the \ac{gp} for \ac{mgt} optimization as 
	\begin{eqnarray*}
		&&(b^{k},c^{k}_\mathrm{G})= \arg\min_{b,c}  \mathcal{H}(s),\\
		&&	\begin{aligned}
			\textnormal{s.t.:}~	& \textnormal{Vehicle dynamics and limits,}	 &&\\
			&		\textnormal{Component models,}	 &&\\
			&\begin{cases}
				x(s)\coloneqq x^{k}(s) \\
				\lambda(s)\coloneqq \lambda^{k}(s)\\
			\end{cases} &&\forall s \in [0,S].&&
		\end{aligned}
	\end{eqnarray*}
\end{prob}

\subsection{Discussion}
Since the \ac{cp} and \ac{gp} both minimize lap time, iterating between the two will yield a situation where neither problem can independently yield an improved lap time that is feasible for the original problem. 
Since we dampen our costate trajectories between iterations, the gearshift trajectory converges at this point, which results in convergence of all other trajectories. 
In our extended work, we provide formal proofs showing that at convergence our solutions satisfy necessary conditions for optimality.
The combination of these findings---even if not sufficient---with the quality and consistency of the numerical results provided in Section~\ref{sec:results} below is promising.

\section{Results and Validation}\label{sec:results}
This section presents the results obtained by applying our framework to various transmission types. We first evaluate the performance of each transmission over a racing lap, and then validate the performance of our iterative algorithm.

\subsection{Numerical Results}
\begin{figure}[t]
	\includegraphics{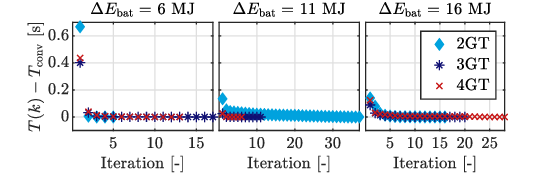} 
	\caption{Convergence of the iterative algorithm, showing the difference between the lap time at each iteration and the final converged lap time $(T_\mathrm{conv})$. Results are shown for multiple battery energy consumption limits.}
	\label{fig:res_basic_conv}
\end{figure}

In this section, we apply our framework to one lap around the \unit[4.2]{km} long Zandvoort race track, using the InMotion LMP3 race car~\cite{InMotion} as a demonstrator. 
We solve Problem~\ref{prob:convex} for the \ac{fgt}- and \ac{cvt}-equipped cars, and apply the iterative algorithm for the cars with a \ac{2gt}, a \ac{3gt} and a \ac{4gt}. 
To realistically compare performance, we include the transmission weight in our simulation. With respect to the \ac{fgt}-equipped car, each added gear in an \ac{mgt} increases the total vehicle weight by \unit[0.37]{\%}, while the \ac{cvt} increases the vehicle weight by \unit[2.6]{\%}.
We parse the problems with YALMIP \cite{Loefberg2004}, using a forward Euler discretization  and a step of \unit[$\Delta s=4$]{m}, and solve with MOSEK \cite{MosekAPS2010} on a laptop with a \unit[2.6]{GHz} processor and \unit[16]{GB} RAM.
Problem~\ref{prob:convex} takes on average \unit[1.2]{s} to solve for the \ac{fgt} and \unit[1.4]{s} for the \ac{cvt}, while the iterative algorithm converges in about \unit[63]{s} on average. 
Fig.~\ref{fig:res_basic_conv} shows that the algorithm gets close to the final converged lap time within a few iterations, while we observe no conclusive correlation between the number of gears and the amount of iterations required for convergence.

\begin{figure}[t]
 	\includegraphics{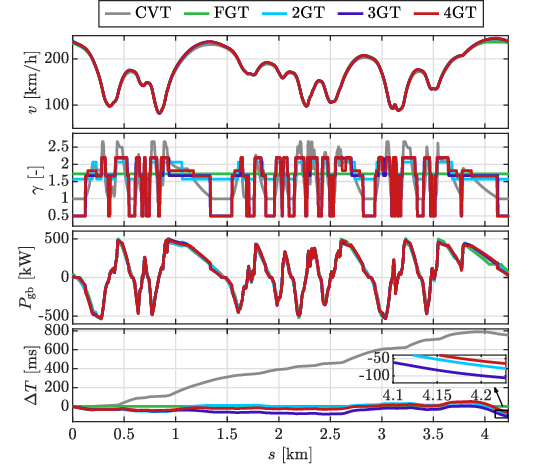} 
 	\caption{Velocity, gear ratio, and transmission output power trajectories for each car, in battery-limited operation. To compare, we also show the relative lap time of each car with respect to the \ac{fgt}-equipped car.}
 	\label{fig:res_basic}
 \end{figure}

\begin{figure}[t]
 	\includegraphics{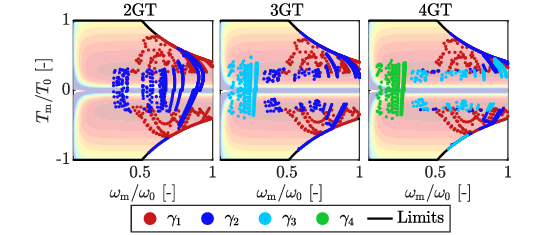}
 	\caption{Operating points of the \ac{em} for each \ac{mgt}-equipped car. }
 	\label{fig:res_gearmap}
 \end{figure}

 Fig.~\ref{fig:res_basic} and Fig.~\ref{fig:res_gearmap} present the results for a case where the vehicle is significantly energy-limited. As can be observed, the \ac{cvt}-equipped car performs considerably worse than the others, due to its lower internal efficiency and higher weight. The \ac{2gt}-car, instead, outperforms the \ac{fgt}-car due to the increased \ac{em} efficiency, which allows it to accelerate faster and regeneratively brake both later and more aggressively. The \ac{3gt}-car contains an added gear with a very low ratio, which allows for a high \ac{em} efficiency at high speed but low power sections. In combination with a higher first gear ratio, this allows the \ac{3gt}-car to accelerate faster out of corners compared to the \ac{2gt}-car, yielding even faster lap times. However, the \ac{4gt}-car cannot provide enough additional performance with its added gear to compensate for its higher weight, rendering it marginally slower than the \ac{2gt}-car. 
 In conclusion, using a \ac{3gt} yields the best trade-off between control freedom and weight for the case presented here.

\subsection{Validation}
In order to validate the efficacy of the iterative algorithm, we solve the \ac{miocp} directly using the MOSEK mixed-integer second-order cone programming solver, which employs a branch-and-bound algorithm to solve the problem with global optimality guarantees. 
Due to the complex nature of our problem, the mixed-integer solver can only solve for a discretization of \unit[$N_\mathrm{steps}=22$]{steps} in less than \unit[$10$]{hours}, requiring exponentially more time for each added step. 
Since the original problem contains \unit[$N_\mathrm{steps}=1058$]{steps}, this means that the mixed-integer solver is only able to solve a section with a length of about \unit[2]{\%} of the lap within \unit[$10$]{hours}. 
Therefore, we validate the iterative algorithm by using a coarser discretization, and by optimizing only a small section of the track, containing a braking zone, a corner, and an acceleration zone. 
To exemplify the computational complexity of our problem, we present the computation times for the mixed-integer solver and iterative algorithm for a varying number of discretization steps in Table~\ref{tab:times_misocp}. 
As can be observed, the mixed-integer algorithm soon becomes intractable, while the iterative algorithm obtains virtually the same section times in much less computation time. 
Only for one simulation is the section time obtained by the iterative algorithm \unit[0.3]{milliseconds} slower, probably due to a poor initial guess. Overall, this validation shows that the iterative algorithm can provide very promising solutions.
 
\begin{table}[t]
	\centering
	\caption{Computation time comparison between a mixed-integer solver and the iterative algorithm on a short track section.}
	\label{tab:times_misocp}
	\begin{tabular}{r|l|l|l|l|l}
		& \multicolumn{2}{c|}{\textbf{\textbf{Mixed-integer}}}   & \multicolumn{2}{c|}{\textbf{\textbf{Iterative}}}  & \\ 
		& \multirow{2}{7.5mm}{Solving time} & \multirow{2}{6mm}{Section time} & 
		\multirow{2}{7.5mm}{Solving time} & \multirow{2}{6mm}{Section time} & \multirow{2}{14.5mm}{Section time difference}\\${N_\mathrm{steps}}$ &&&&&\\ \hline
		12   & \unit[43]{s}    & \unit[3.5946]{s} & \unit[2]{s} & \unit[3.5946]{s} & \unit[0.0]{ms}\\ 
		14   & \unit[120]{s}   & \unit[3.4959]{s} & \unit[2]{s} & \unit[3.4959]{s} & \unit[0.0]{ms}\\ 
		16   & \unit[575]{s}   & \unit[3.4256]{s} & \unit[2]{s} & \unit[3.4259]{s} & \unit[0.3]{ms}\\ 
		18   & \unit[1940]{s}  & \unit[3.3403]{s} & \unit[2]{s} & \unit[3.3403]{s} & \unit[0.0]{ms}\\ 
		20   & \unit[8153]{s}  & \unit[3.4759]{s} & \unit[2]{s} & \unit[3.4759]{s} & \unit[0.0]{ms}\\ 
		22   & \unit[34915]{s} & \unit[3.4302]{s} & \unit[2]{s} & \unit[3.4302]{s} & \unit[0.0]{ms}\\ 
	\end{tabular}
\end{table}

\section{Conclusion}\label{sec:conclusion}
In this paper, we presented a framework to optimize the design and control of an electric race car, considering a \acf{cvt}, a \acf{fgt} and a \acf{mgt}.
To this end, we developed an iterative algorithm to efficiently handle the mixed-integer \ac{mgt} gearshifts, combining convex optimization and \acf{pmp}.
We demonstrated that our algorithm satisfies necessary conditions for optimality upon convergence, and corroborated this with numerical results showing convergence to promising solutions in terms of optimality.
Finally, we studied the performance of the various transmissions on the Zandvoort race track, where we observed that an \ac{mgt} can balance the individual advantages of both an \ac{fgt} and a \ac{cvt}, by delivering significant control over the \acf{em} operating range at a low cost in terms of transmission weight and efficiency loss.
Interestingly, we also noted that adding one gear too many can be detrimental to the lap time, highlighting the importance of carefully choosing the right transmission technology and design for specific car and track requirements.
In a future extended work we will provide detailed models, formal proofs, and design studies over a range of battery energy levels and \ac{em} scales.

\section*{Acknowledgments}
\noindent We thank Ir. J. van Kampen, Ir. O. Borsboom and Dr. I. New for their comments and advice. 

\bibliographystyle{IEEEtran}
\bibliography{bibliography}             

\end{document}